\documentclass[ajl]{emulateapj}

\usepackage{graphicx,epsfig,natbib,color}
\usepackage{lscape}
\usepackage{graphicx}
\usepackage{epsfig}
\usepackage{natbib}

\slugcomment{Accepted for publication in ApJ Letters on April 28th}

\begin{document}
\title{Investigating Two Successive Flux Rope Eruptions In A Solar Active Region}

\author{X. Cheng\altaffilmark{1,2}, J. Zhang\altaffilmark{1,3}, M. D. Ding\altaffilmark{1,2},
O. Olmedo\altaffilmark{4}, X. D. Sun\altaffilmark{5}, Y. Guo\altaffilmark{1,2} \& Y. Liu\altaffilmark{6}}

\affil{$^1$ School of Astronomy and Space Science, Nanjing University, Nanjing 210093, China}\email{xincheng@nju.edu.cn}
\affil{$^2$ Key Laboratory for Modern Astronomy and Astrophysics (Nanjing University), Ministry of Education, Nanjing 210093, China}
\affil{$^3$ School of Physics, Astronomy and Computational Sciences, George Mason University, Fairfax, VA 22030, USA}
\affil{$^4$ NRC, Naval Research Laboratory, Washington, DC 20375, USA}
\affil{$^5$ W. W. Hansen Experimental Physics Laboratory, Stanford University, Stanford, CA 94305, USA}
\affil{$^6$ State Key Laboratory of Space Weather, National Space Science Center, Chinese Academy of Sciences, Beijing, China}

\begin{abstract}

We investigate two successive flux rope (FR1 and FR2) eruptions resulting in two coronal mass ejections (CMEs) on 2012 January 23. Both FRs appeared as an EUV channel structure in the images of high temperature passbands of the Atmospheric Imaging Assembly prior to the CME eruption. Through fitting their height evolution with a function consisting of linear and exponential components, we determine the onset time of the FR impulsive acceleration with high temporal accuracy for the first time. Using this onset time, we divide the evolution of the FRs in the low corona into two phases: a slow rise phase and an impulsive acceleration phase. In the slow rise phase of the FR1, the appearance of sporadic EUV and UV brightening and the strong shearing along the polarity inverse line indicates that the quasi-separatrix-layer reconnection likely initiates the slow rise. On the other hand for the FR2, we mainly contribute its slow rise to the FR1 eruption, which partially opened the overlying field and thus decreased the magnetic restriction. At the onset of the impulsive acceleration phase, the FR1 (FR2) reaches the critical height of 84.4$\pm$11.2 Mm (86.2$\pm$13.0 Mm) where the decline of the overlying field with height is fast enough to trigger the torus instability. After a very short interval ($\sim$2 minutes), the flare emission began to enhance. These results reveal the compound activity involving multiple magnetic FRs and further suggest that the ideal torus instability probably plays the essential role of initiating the impulsive acceleration of CMEs.

\end{abstract}

Online-only material: animations, color figures

\section{Introduction}
Coronal mass ejections (CMEs) are the most spectacular eruptive phenomena in our solar system. They are able to release a large quantity of plasma and magnetic flux into the heliosphere and severely disturb the space environment around the Earth \citep[e.g.,][]{gosling93}. Over the last 20 years, although the solar community has made a considerable progress in many aspects of understanding CMEs, the important issue of how CMEs are initiated remains elusive \citep{chen11_review,schmieder12}. In this Letter, we report the compound eruption activity involving two inter-connected flux ropes (FRs; magnetic field lines wound around each other) in the same active region, and then elucidate their initiation mechanism.

In terms of whether involving magnetic reconnection, existing initiation models can be divided into two groups; one group are reconnection-based models including tether-cutting reconnection \citep{moore01} and breakout reconnection \citep{antiochos99,chen00,karpen12}, and the other group are FR-based ideal magnetohydrodynamics (MHD) models, involving catastrophic loss-of-equilibrium \citep{forbes91,isenberg93}, kink instability \citep{torok04}, and torus instability \citep{kliem06}. In the tether-cutting (breakout) model, key mechanism solely concerns the magnetic reconnection in the CME core (overlying) field region that increases upward magnetic pressure (reduce the overlying tension), thus initiating the explosive eruption of CMEs. Differing from the reconnection models, \cite{forbes91} and \cite{isenberg93} showed that the FR will lose equilibrium in an ideal MHD process if the photospheric sources of the overlying field converge toward each other. In addition to this catastrophic loss of equilibrium, kink or torus instability is also capable of initiating the CME explosive eruption. The ideal kink instability develops if the average twist of the FR is greater than a threshold \citep[e.g., 3.5$\pi$;][]{torok04}. The torus instability takes place if the restoring force caused by overlying field decreases faster than the outward directed Lorenz self-force as the flux rope expands outward \citep{kliem06,olmedo10}. 

With the development of these theoretical models, validating and distinguishing them observationally becomes a matter of great necessity. We here investigate in detail the initiation of a compound CME activity originating from two successive FR eruptions observed by the Atmospheric Imaging Assembly \citep[AIA;][]{lemen12} on board \textit{Solar Dynamics Observatory} (\textit{SDO}). We find that the slow rise of the first CME is most likely due to the quasi-separatrix-layer (QSL) reconnection in the low corona; while the slow rise of the second one results from the partial opening of the overlying field by the first CME. However, for their initiation of the impulsive acceleration, we contribute the mechanism to be the ideal torus instability. Data and results are presented in Section 2 and 3, respectively. Summary and discussions are given in Section 4.

\section{Instruments}
The data we used are mainly from the AIA, as well as the Helioseismic and Magnetic Imager \citep[HMI;][]{schou12} on broad \textit{SDO}. AIA provides the EUV images of the solar corona with the temporal cadence of 12 seconds, the pixel size of 0.6\arcsec, and the field of view (FOV) of 1.3$R_\odot$. The HMI provides the vector magnetic field of the solar photosphere with almost the same pixel size and FOV as AIA but the temporal cadence of 12 minutes. The Large Angle and Spectrometric Coronagraph \citep[LASCO;][]{brueckner95} on board \emph{SOHO} and the Sun--Earth Connection Coronal and Heliospheric Investigation \citep[SECCHI;][]{howard08} on board \emph{STEREO} observe the CMEs in white light. \textit{GOES} X-ray data reveal the SXR (SXR) 1--8 {\AA} flux of CME-associated flares.

\section{Results}

\subsection{Kinematics of CMEs}
\begin{figure*} %%%%%%%%%%%%%%%%%% FIGURE 1
      \vspace{-0.0\textwidth}    % Shift back to the panel bottom
      \centerline{\hspace*{0.00\textwidth}
      \includegraphics[width=1.0\textwidth,clip=]{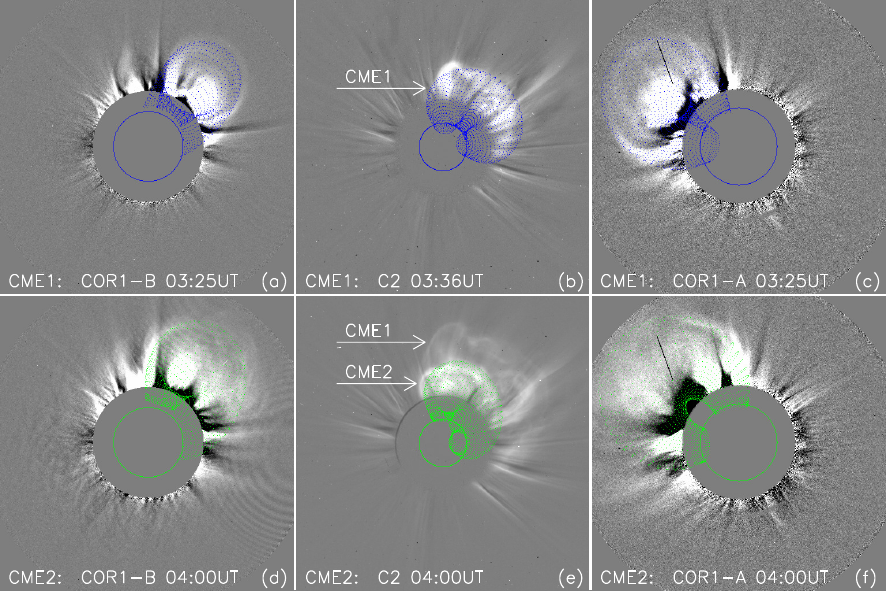}
      }
\caption{ White light coronagraph images and GCS FR (blue and green lines) modeling of the CME1 and CME2 on 2012 January 23. Blue and green circles denote the location of the solar limb.} \label{f1}
\end{figure*}

On 2012 January 23, there are two CMEs (CME1 and CME2) that appeared successively in the FOVs of LASCO/C2 and SECCHI/COR1, finally manifesting as a compound CME in the FOV of LASCO/C3 \citep[also see;][]{joshi13,shen13,liuy13}. The selected snapshots are displayed in Figure \ref{f1}. Here, we take advantage of the graduated cylindrical shell (GCS) model developed by \citet{thernisien06} to reconstruct the three-dimensional (3D) morphology of the CMEs and obtain the true heights of the CMEs. The forward fitting model includes three positional parameters and three dimensional parameters: the Carrington longitude ($\phi$) and latitude ($\theta$), the tilt angle ($\gamma$) with respect to the equator, the height ($h$) and aspect ratio ($\kappa$) of the FR, and the half-angle ($\alpha$) between the two FR legs. For the CME1 at $\sim$03:30 UT, the fitting parameters are $\phi$=219$^\circ$, $\theta$=25$^\circ$, $\gamma$=--49$^\circ$, $h=4.1$ $R_\sun$, $\kappa$=0.4, and $\alpha$=32$^\circ$; for the CME2 at $\sim$04:00 UT, $\phi$=210$^\circ$, $\theta$=32$^\circ$, $\gamma$=--71$^\circ$, $h=4.2$ $R_\sun$, $\kappa$=0.4, and $\alpha$=40$^\circ$. The corresponding wireframe rendering are shown in Figure \ref{f1}.

\begin{figure*} %%%%%%%%%%%%%%%%%% FIGURE 2
     \vspace{-0.0\textwidth}    % Shift back to the panel bottom
     \centerline{\hspace*{0.00\textwidth}
               \includegraphics[width=1.0\textwidth,clip=]{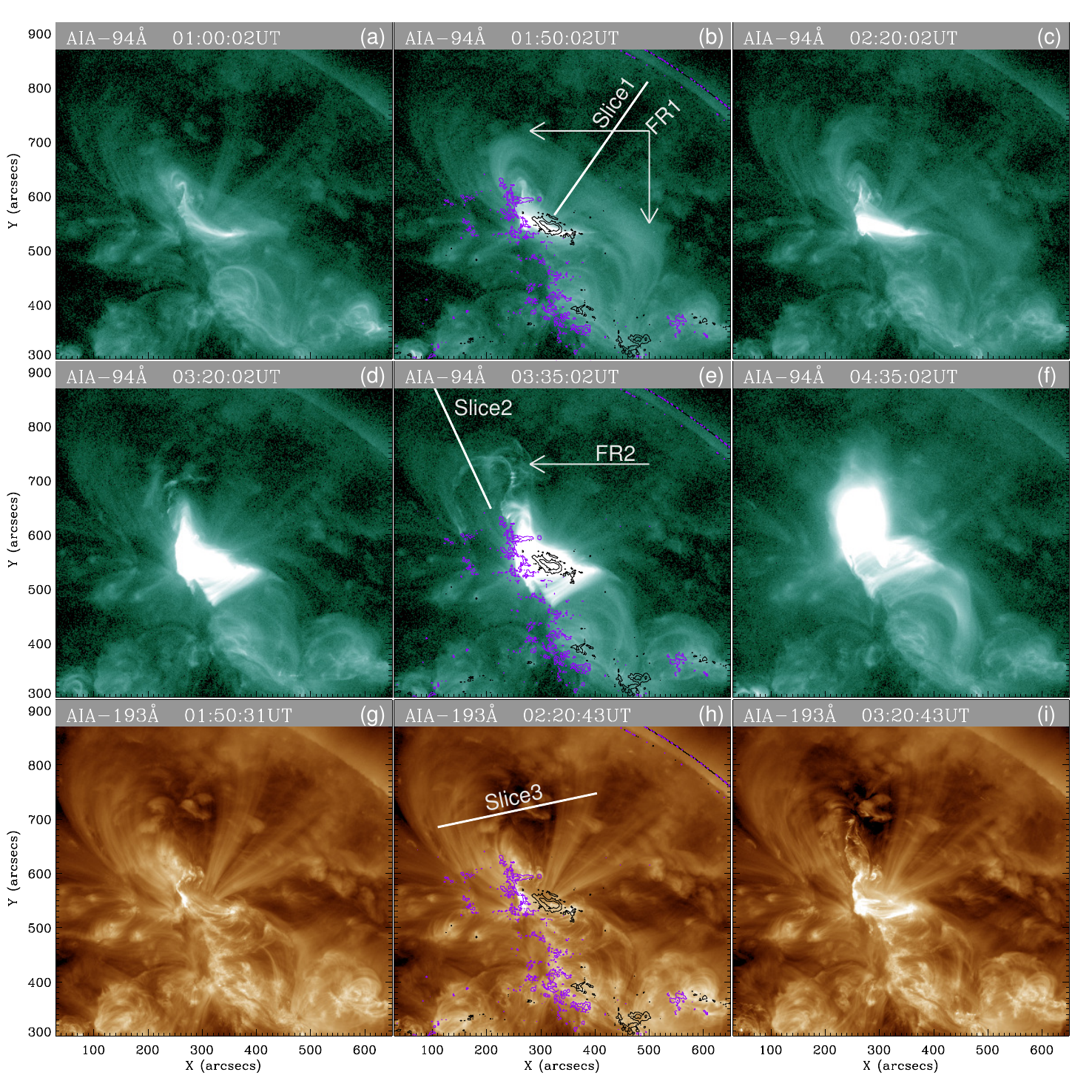}
               }
\caption{(a)--(f) AIA 94 {\AA} ($\sim$7 MK) images of the CME1 and CME2 HCS. (g)--(i) AIA 193 {\AA} ($\sim$1 MK)  images of the overlying field. In panels (b), (e), and (h), overlaid contours show the positive (purple) and negative (black) polarity of magnetic field. The two arrows point out the two FRs. The three oblique solid lines indicate the orientations of the slice1, slice2, and slice3.} \label{f2}

(Animations of this figure are available in the online journal.)
\end{figure*}

\begin{figure*} %%%%%%%%%%%%%%%%%% FIGURE 3
     \centerline{\hspace*{0.0\textwidth}
               \includegraphics[width=0.55\textwidth,clip=]{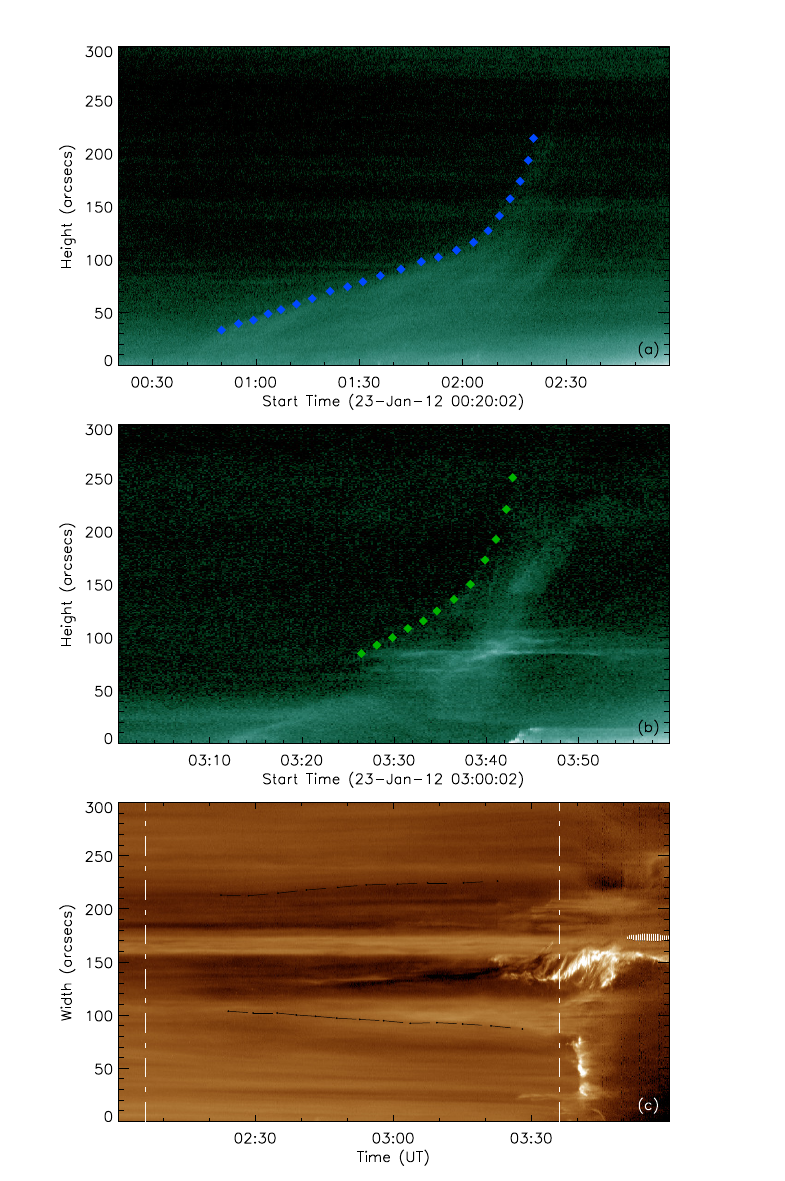}
               \hspace*{-0.1\textwidth}
               \includegraphics[width=0.55\textwidth,clip=]{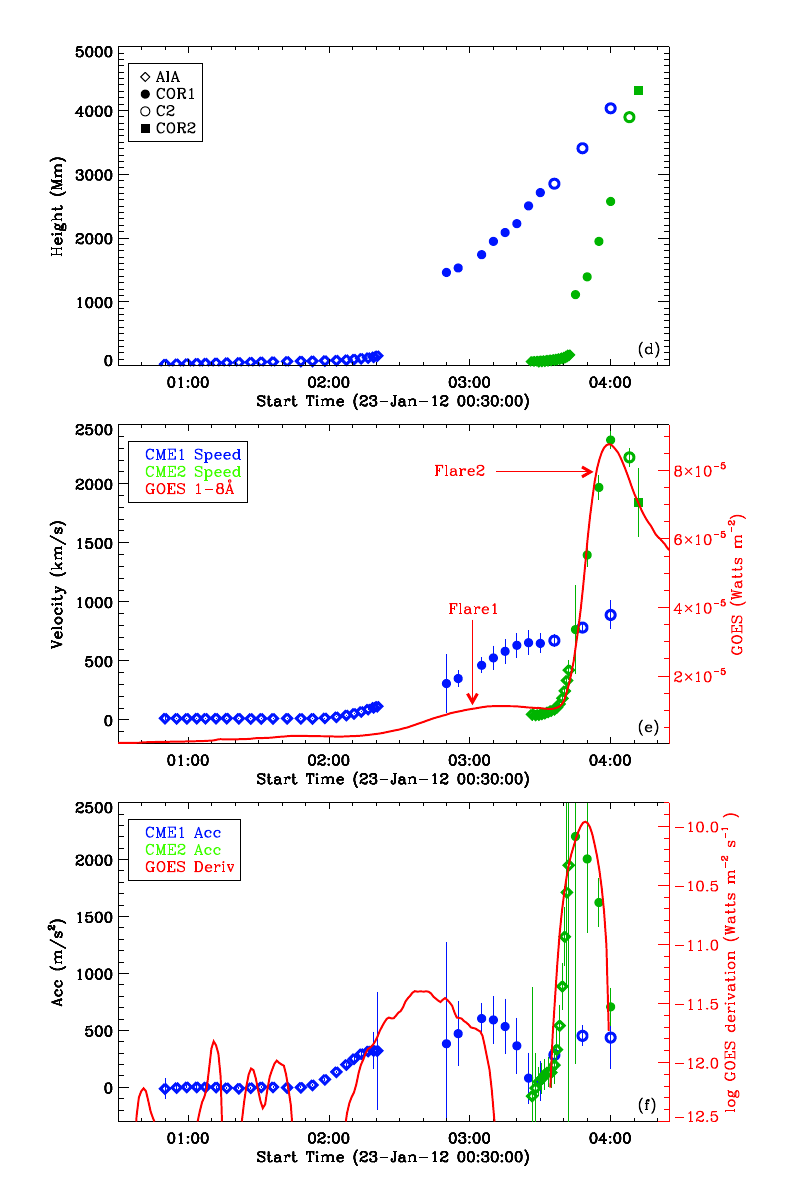}
               }
\caption{(a)--(c) Stack plots of the intensity along the slice1, slice2, and slice3 as shown in Figure \ref{f2}. The diamonds in panel (a) and (b) indicate the height-time measurements of the CME HCSs, two vertical lines in panel (c) denote the onsets of the FR1 and FR2, two inclined lines show the expansion of the overlying field. (d)--(f) Temporal evolution of the heights, velocities, and accelerations of the CME1 and CME2. The GOES SXR 1--8 {\AA} flux and resulting time derivation are shown in panels (e) and (f), respectively. Note that the acceleration of the FR1 is tripled to compare with the time derivation of the flare1.} \label{f3}
\end{figure*}

The FR of CMEs have been reported to appear as an EUV hot channel structure (HCS) \citep{zhang12,cheng12_dem,cheng13_driver,li13} and/or bubble \citep{cheng11_fluxrope,suyang12,patsourakos13} in the AIA images. For the two CMEs studied we inspect the AIA images and find that the FR1 only appeared in 131 {\AA} and 94 {\AA} passbands (Figure \ref{f2}(b)) but not in any other ones, indicating that it should be hot\citep[$\ge$7MK;][]{odwyer10,reeves11,cheng11_fluxrope}. On the other hand, the FR2 is visible in all AIA passbands, although most noticeable in 131 {\AA} and 94 {\AA} images (Figure \ref{f2}(e)). The mixture of hot and cold
plasmas within the FR2 probably is attributed to the presence of filament material \citep[e.g.,][]{cheng12_dem}. More detailed evolution of the two HCSs please refers to the attached movies. In order to clearly display the rising motion of the HCSs, as well as the expansion of the overlying field, we make two slices (slice1 and slice2 in Figure \ref{f2}(b) and (e)) along the HCS rising directions and one slice (slice3 in Figure \ref{f2}(h)) along the overlying field expanding direction. The time evolution of the HCSs and the overlying field along these slices are shown in the slice-time plots in Figure \ref{f3}(a)--(c). Through these time-stacking plots, we measure the height of the two HCSs with time (diamonds in Figure \ref{f3}(a) and (b)). The height--time data are plotted in Figure \ref{f3}(d). From Figure \ref{f2}(g)--(i), as well as the stack plot of the slice3 (as shown by two inclined lines in Figure \ref{f3}(c)), we notice the slow expansion of the overlying field after the FR1 eruption, which most probably leads to the initial rise motion of the FR2.

With the height-time data, we further calculate the velocity and acceleration. In order to reduce the uncertainty in the height measurement, a cubic spline smoothing method is used to smooth the height \citep[also see,][]{patsourakos10_genesis,cheng13_driver}. Then a piece-wise numerical derivative method is applied to calculate the velocity \citep[e.g.,][]{zhang01,zhang04,cheng10_buildup}. The deduced velocity--time profile is displayed in Figure \ref{f3}(e). The uncertainty in the velocity is mainly from the uncertainty in the height measurement (to be 4 pixels; 1700 km for AIA observations and 44,000 km for GCS fitting). Similarly, we derive the CME acceleration and resulting uncertainty, as shown in Figure \ref{f3}(f). Note that all heights refer to the top of the CME FRs from the solar surface.

From Figure \ref{f3}(e) and (f), one can find that the kinematical evolution of the CMEs are tightly associated with the emission variation of associated flares. The velocity evolutions of the CME1 and CME2 are very consistent in time with the SXR 1--8 {\AA} flux profiles associated with flare1 and flare2. The acceleration of the CME1 and CME2 grows in step with the derivative of the SXR profile. The close temporal correlation between the CME kinematics and the flare emission implies that CMEs and associated flares are not two independent eruption phenomena but only two distinct manifestations of the same eruption process \citep{linjun00,priest02,zhang06,temmer10}.

\subsection{Onset of Impulsive Acceleration of CMEs}
In order to investigate the initiation of the CMEs, we first determine the onset of the impulsive acceleration. Assuming that the height evolution of the CME FR in the low corona follows a function $h(t)=c_{0}e^{(t-t_{0})/\tau}+c_{1}(t-t_{0})+c_{2}$, where $h(t)$ is height, $t$ is time, $\tau, t_{0}, c_{0}, c_{1}, c_{2}$ are five free coefficients. The model includes two components: the linear and the exponential, which correspond to the slow rise phase characterized by a constant velocity and the initial impulsive acceleration phase characterized by an exponentially increase of velocity, respectively. The model is reasonable due to the fact that the velocity of the CMEs increases rapidly once the impulsive acceleration is triggered whether by the flare reconnection \citep[e.g.,][]{antiochos99,moore01,karpen12} or by the torus instability \citep[e.g.,][]{torok05,olmedo10}. The fitting to the data is achieved by the ``mpfit.pro" routine. With the fitting parameters, the onset of the CMEs is defined as a time where the exponential velocity is equal to the linear velocity ($t_{\rm onset} $=$\tau \ln (c_{1} \tau /c_{0})+t_{0}$). Accordingly, the height at the onset time corresponds to the critical height $h({t_{\rm onset}})$ of the eruption. 

\begin{figure*} %%%%%%%%%%%%%%%%%% FIGURE 4
     \centerline{\hspace*{0.02\textwidth}
               \includegraphics[width=.45\textwidth,clip=]{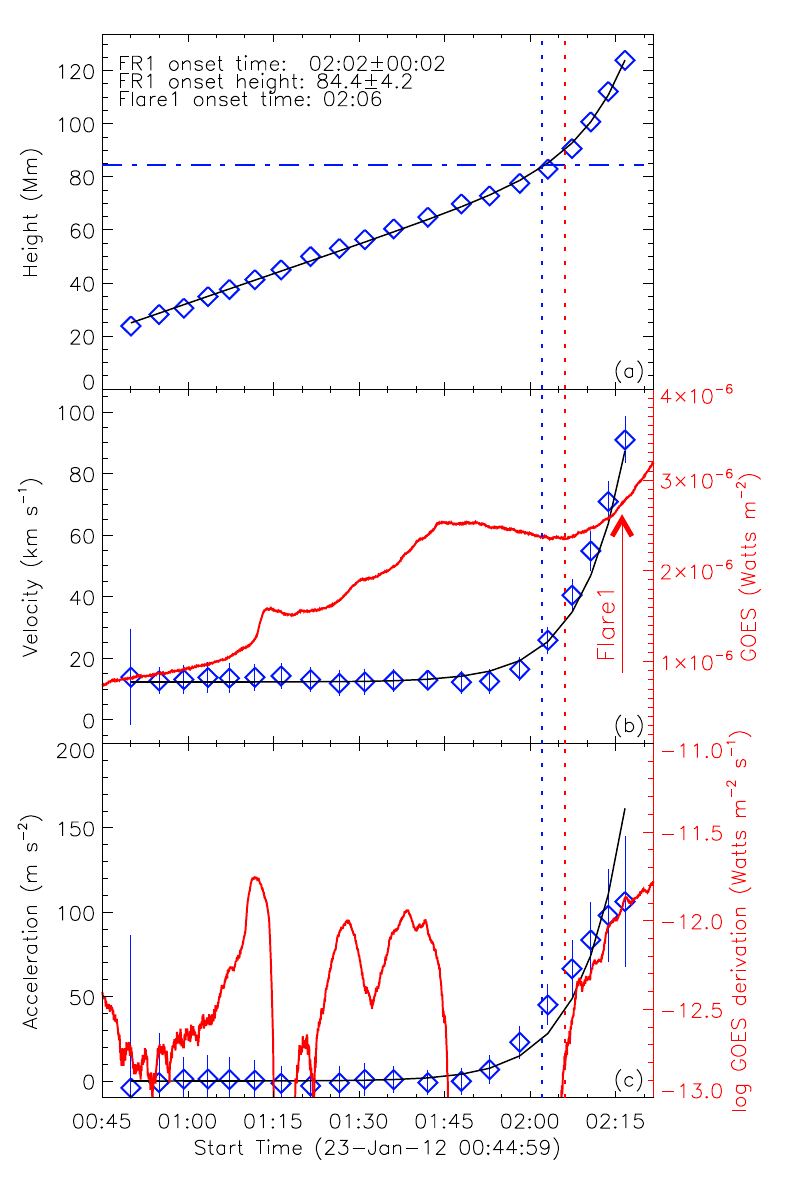}
                       \hspace*{-0.02\textwidth}
               \includegraphics[width=.45\textwidth,clip=]{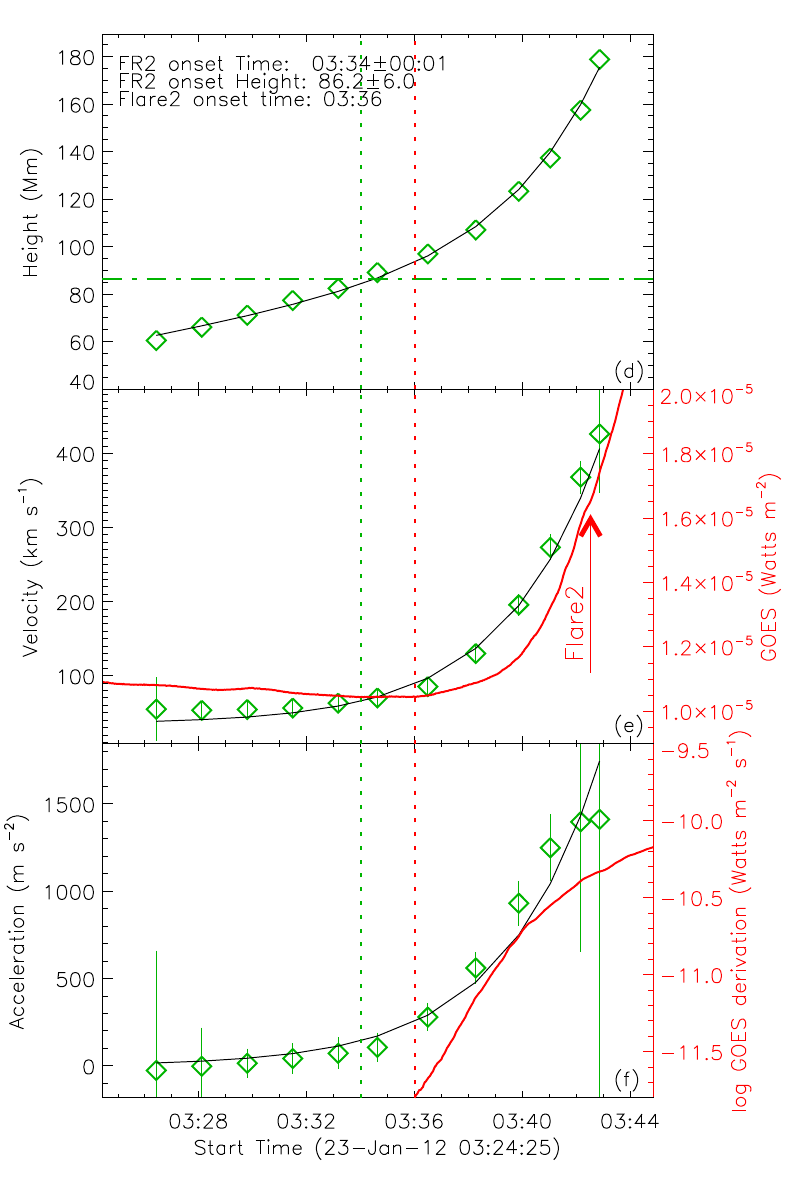}
               }
\caption{(a)--(c) Temporal evolution of the height, velocity, and acceleration of the CME1 HCS in the FOV of AIA. The black solid lines show the model fitting to measured height, velocity, and acceleration, respectively. The red solid lines show the GOES SXR 1--8 {\AA} flux and resulting time derivation. The vertical blue (horizontal) line points out the onset time (height) of the FR1. The vertical red line indicates the onset time of the flare1. (d)--(f) Same as (a)--(c) but for the CME2 HCS. The vertical green (horizontal) line denotes the onset time (height) of the FR2. The vertical red line shows the onset time of the flare2.} \label{f4}
\end{figure*}

We further use 100 Monte Carlo (MC) simulations to estimate the uncertainties of our results. For each MC realization, the measured heights are first perturbed randomly by an amount of $\delta$ within a sigma equal to the error of the height, and then re-fitting the model to the heights. The final onset time and onset height are the averages of 100 onset times and onset heights derived by 100 MC realizations. The corresponding uncertainties are triple the standard deviations (3$\sigma$) of the MC fitting. The fitting results are shown in Figure \ref{f4}. For the CME1, we determine the onset at 02:02 UT with the error of 2 minutes; the onset height is 84.4$\pm$4.2 Mm. For the CME2, the onset is 03:34 UT with the error of 1 minute; the onset height is 86.2$\pm$6.0 Mm. Moreover, the uncertainty of the reference point of measuring height is estimated to be $\sim$7.0 Mm (10$''$). Therefore, the final uncertainties of the onset heights for the CME1 and CME2 are $\sim$11.2 Mm and 13.0 Mm, respectively.

From Figure \ref{f4}(c) and (f), one can find that the onset of the CMEs is earlier than that of associated flares by a few minutes. Here we define the onset of flares as a time where the derivative of the SXR flux becomes positive and begins to increase successively. For CME1 and CME2, the leading times, i.e., the onset of CME impulsive acceleration relative to the onset of the flare impulsive phase, are about 2 minutes. Actually, for the flare2 the onset time in \emph{GOES} record is 03:38 UT; the leading time would be of $\sim$4 minutes if this onset time is adopted. The results imply that the impulsive acceleration onset of the CME FRs is most probably caused by ideal MHD instability rather than by the flare reconnection. In addition, we note that the impulsive acceleration of the FRs should occur earlier than that we determined if the projection effect for the heights in the AIA FOV is taken into account. The result should be strengthened because it is even earlier for the FRs to ascend to the critical height of the torus instability.

\subsection{Magnetic Properties of CMEs}
\begin{figure*} %%%%%%%%%%%%%%%%%% FIGURE 5
     \centerline{\hspace*{0.00\textwidth}
               \includegraphics[width=1.0\textwidth,clip=]{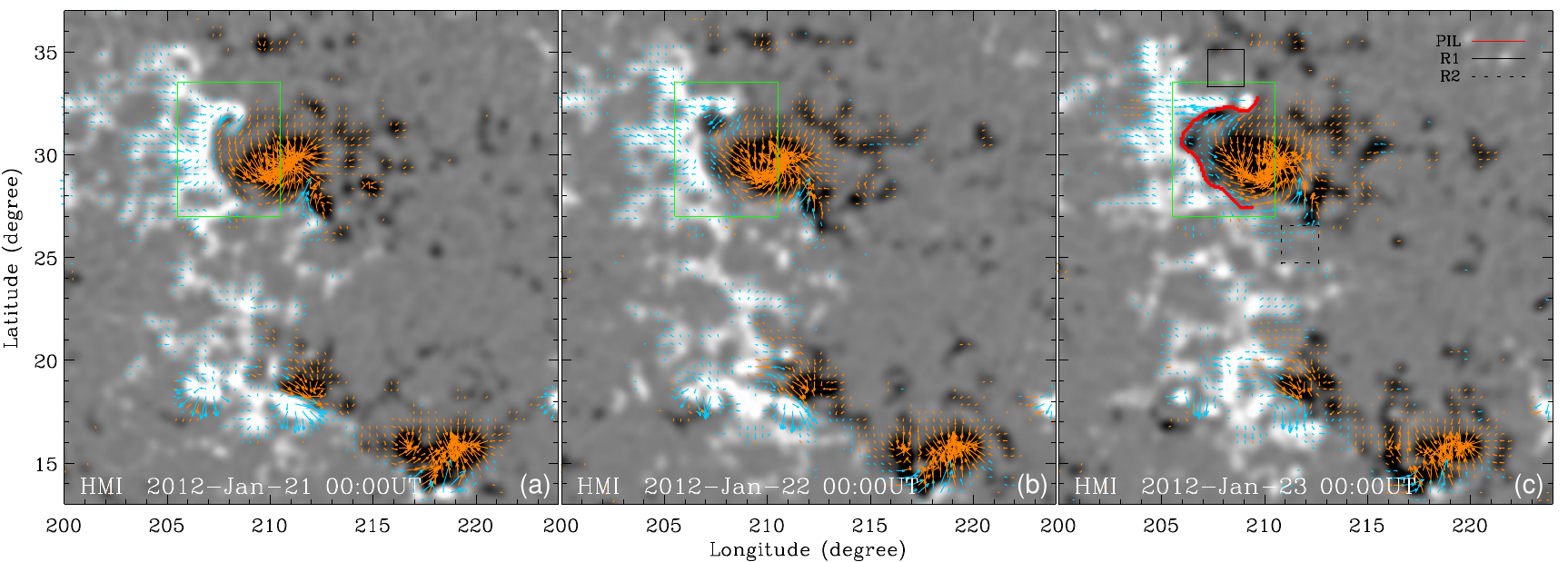}}
     \centerline{\hspace*{0.00\textwidth}
               \includegraphics[width=1.0\textwidth,clip=]{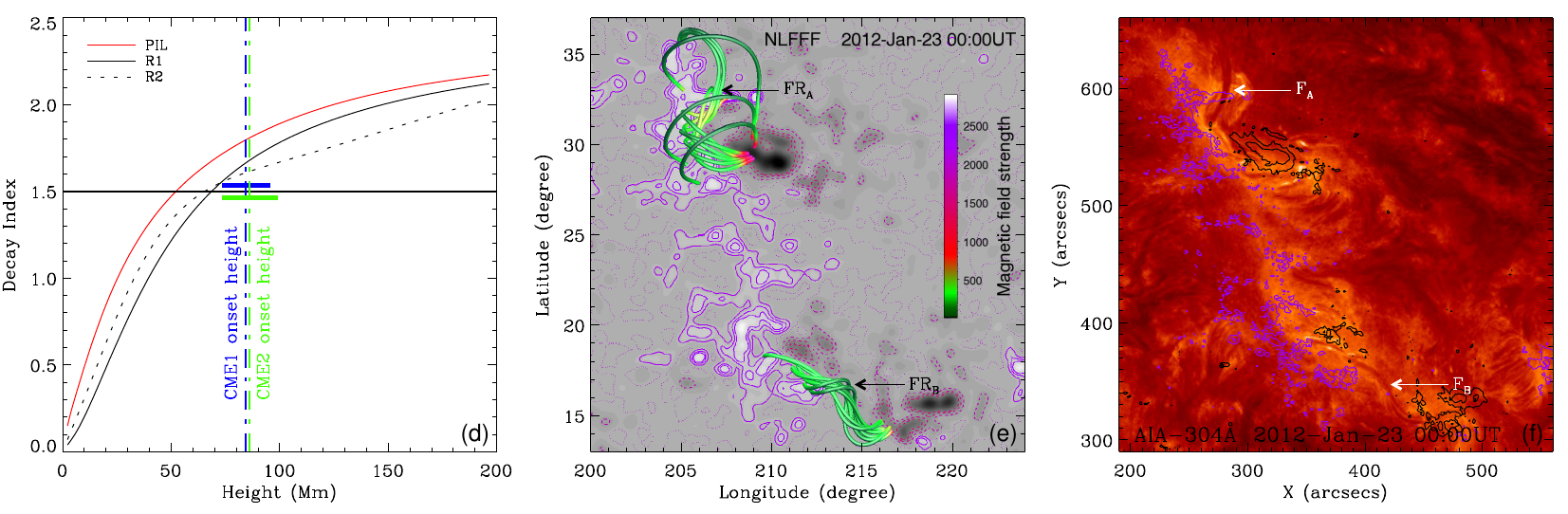}}
\caption{(a)--(c) HMI vector magnetograms with the 180{\degr} ambiguity removed. The background images are the processed vertical magnetograms. The green boxes indicate the main AR that the two HCSs origin from. The red curve in panel (c) shows the main PIL. The two black boxes point out two regions (R1 and R2) near the main PIL. (d) Distributions of the horizontal magnetic field decay index with height over the main PIL, R1 and R2. The two vertical lines display the onset heights of the CME1 and CME2, respectively. The widths of the blue and green bars show the uncertainties in the onset heights. (e) Top view of the extrapolated 3D magnetic field configuration. The background image is the processed vertical magnetogram overlaid with the contours. The two arrows indicate the FR$\rm_{A}$ and FR$\rm_{B}$. (f) AIA 304 {\AA} ($\sim$0.05 MK) image overlaid with the contours of line-of-sight magnetic field, the purple (black) means the positive (negative) polarity. The two arrows denote the filament A (F$\rm_{A}$) and filament B (F$\rm_{B}$).} \label{f5}

(An animation of this figure is available in the online journal.)
\end{figure*}

Previous theoretical works have proved that the occurrence of torus instability of a FR depends on the decay index of the background magnetic field $B$ with height $h$ \citep[$n$=$-d \ln B/d \ln h$;][]{kliem06,isenberg07,olmedo10,demoulin10}. The decay index is computed from the potential field model based on the vertical component of vector magnetograms provided by the HMI (Figure \ref{f5}(a)--(c)). The distributions of the average decay index with height over three regions (the main polarity inverse line (PIL), two rectangle regions near the main PIL; Figure \ref{f5}(c)) are shown in Figure \ref{f5}(d). One can find that at the onset of the CME impulsive eruption, the FR1 (FR2) reached the height of 84.4$\pm$11.2 Mm (86.2$\pm$13.0 Mm) where the decay index is 1.7$\pm$0.1, which is slightly larger than the nominal critical value $\sim$1.5 of the torus instability occurrence \citep{torok05}. This result supports the theory of the torus instability as the trigger of the impulsive acceleration of CMEs. Note that here the decay index is calculated from the horizontal component of 3D magnetic field because the vertical component does not have a role in constraining the FR. Also note that, in torus instability models, the field induced by the current inside the FR and the background field constraining the FR constitute the total magnetic field, which usually can be modeled by the nonlinear force-free field (NLFFF) \citep[e.g.,][]{guo10_filament,sun12}. Considering the fact that it is difficult to separate the background field from the FR in the NLFFF model, the potential field thus is used to be an approximation of the background field \citep[also see,][]{fan07,aulanier10,cheng13_driver}.

Based on the HMI vector data at 00:00 UT on January 23 (Figure \ref{f5}(c)), we extrapolate the 3D NLFFF using the optimization algorithm in cartesian coordinate \citep{wheatland00,wiegelmann04}. The selected field lines are displayed in Figure \ref{f5}(e). One can find that the extrapolation suggests there exist two FR structures (marked as FR$_{\rm A}$ and FR$_{\rm B}$), both of which have an associated filament visible in the AIA 304 {\AA} passband (Figure \ref{f5}(f)). The FR$_{\rm A}$ finally erupted as the CME2, while the FR$_{\rm B}$ always stayed there during the whole eruption process. However for the FR1 discussed later, we are not able to reconstruct a corresponding FR even resorting to the optimization algorithm in spherical geometry with a larger FOV \citep[e.g.,][]{guo12_spher}. This may not be surprising, since the FR1 as observed in AIA is a larger and higher structure, which can not be well modeled with the current extrapolation technique.

\section{Summary and Discussions}
In this Letter, we report the initiation process of a compound CME activity consisting of two successfully erupted FRs. We find that the kinematics of the FRs in the low corona have two phases: a slow rise phase and an impulsive acceleration phase, which can be fitted very well by a model consisting of a linear component and an exponential component. 

In the slow rise phase of the FR1, we inspect the AIA images in all EUV and UV passbands and find that some brightenings spread sporadically along the PIL under the FR1 and at its two footpoints as well. It indicates that magnetic reconnection probably take place inside or around the FR1. The reconnection is most likely to be at the QSL surrounding the FR in the low corona \citep{aulanier10}. The QSL reconnection is too weak to produce nonthermal particles. It is different from the reconnection process occurring during the flare, in terms of both intensity and location. With this QSL reconnection taking place, the FR1 is allowed to grow and rise slowly. However for the FR2, the driving mechanism of the slow rise phase might be different from that of FR1. Due to the eruption of the FR1, the overlying field is opened partially, leading to the expansion of the ambient magnetic field. The expansion decreases the downward magnetic tension, causing the slow rise of the FR2 \citep{torok11,lynch13}. 

Through 3D NLFFF extrapolation, it is indicated that the FR2 exists prior to the slow rise phase for a long time. While for the FR1, we do not know whether it exists or not prior to the slow rise. It may be formed locally through the QSL reconnection in the slow rise phase \citep[e.g.,][]{aulanier10,liur10} or there has existed a nascent FR, and the QSL reconnection only plays a role in enhancing the poloidal flux of the FR1. 

In the end of the slow rise phase, the FR1 (FR2) reaches the height of 84.4$\pm$11.2 Mm (86.2$\pm$13 Mm), where the decrease of the background magnetic field with height is fast enough such that the torus instability takes place, thus triggering the impulsive acceleration. It is worth mentioning that we do not find evidence of the kink instability from the observations.

As the impulsive acceleration commences, the FR quickly stretches the anchored overlying field upward, and then the magnetic field underneath starts to reconnect impulsively. Such runaway reconnection is the cause of the observed flare rise phase. The time lag between the onset of associated flare and that of torus instability is only a few minutes, probably shorter. Without high cadence observations from AIA, it would be difficult to distinguish the relative timing between flux rope impulsive acceleration and flare reconnection onset.

As the flare starts, the CME acceleration and the flare reconnection are coupled together. The flare reconnection is able to rapidly convert ambient magnetic field into the FR, enhancing the upward Lorentz self-force, thus further accelerating the FR. In response to the escaping of the FR, reduced magnetic pressure would drive a plasma inflow, which in turn causes more ambient magnetic field reconnecting. Therefore, the runaway reconnection after the flare onset is in a positive feedback process that impulsively accelerates the CME and enhances the flare emission. 

%%%%%%%%%%%%%%%%%%%%%%%%%%%%%%%%%%%%%%%%%%%%%
\acknowledgements We thank the referee, T. T{\"o}r{\"o}k, Y. M. Wang, C. L. Shen, R. Liu, K. Liu, and P. F. Chen for their valuable comments and discussions. SDO is a mission of NASA's Living With a Star Program. X.C., M.D.D., and Y.G. are supported by NSFC under grants 10673004, 10828306, and 10933003 and NKBRSF under grant 2011CB811402. J.Z. is supported by NSF grant ATM-0748003, AGS-1156120 and NASA grant NNG05GG19G.

%%%%%%%%%%%%%%%%%--\appendix--%%%%%%%%%%%%%%%%%%%%%%
%%%%%%%%%%%%%%%%%--references--%%%%%%%%%%%%%%%%%%%%%
%\bibliographystyle{apj} 
%\bibliography{reference}

%%%%%%%%%%--FIGUREs--%%%%%%%%%%%%%%%%%%%%%%%%%%%%

%%%%%%%%%%%%%%%%%%%%%%%%%%%%%%%%%%%%%%%%%%%%%

%%%%%%%%%%%%%%%%%%%%%%%%%%%%%%%%%%%%%%%%%%%%%

%%%%%%%%%%%%%%%%%%%%%%%%%%%%%%%%%%%%%%%%%%%%%

%%%%%%%%%%%%%%%%%%%%%%%%%%%%%%%%%%%%%%%%%%%%%

%%%%%%%%%%%%%%%%%%%%%%%%%%%%%%%%%%%%%%%%%%%%%
\end{document}